\documentclass[prc,twocolumn,showpacs,floatfix,nofootinbib,preprintnumbers,superscriptaddress,amsmath,amssymb]{revtex4}
\usepackage{graphicx}
\usepackage{subfigure}
\usepackage{amsmath}
\usepackage{amsfonts}
\usepackage{float}
\usepackage{color}

\newcommand{\UNEDFZERO}{\textsc{unedf0}}

\newcommand{\HOSPHE}{\textsc{hosphe}}

\newcommand{\enm}{E^\text{NM}}
\newcommand{\rhoc}{\rho_\text{c}}

\newcommand{\asym}{a_\text{sym}^\text{NM}}
\newcommand{\lsym}{L_\text{sym}^\text{NM}}

\newcommand{\sub}[1]{\text{\tiny{#1}}}

\begin{document}

\title{Propagation of uncertainties in the Skyrme energy-density-functional model}

\author{Y. Gao}
\affiliation{Department of Physics, P.O. Box 35 (YFL), FI-40014
University of Jyv\"askyl\"a, Finland}

\author{J. Dobaczewski}
\affiliation{Department of Physics, P.O. Box 35 (YFL), FI-40014
University of Jyv\"askyl\"a, Finland}
\affiliation{Institute of Theoretical Physics, Faculty of Physics,
University of Warsaw, ul. Ho\.za 69, PL-00-681 Warsaw, Poland}

\author{M. Kortelainen}
\affiliation{Department of Physics, P.O. Box 35 (YFL), FI-40014
University of Jyv\"askyl\"a, Finland}

\author{J. Toivanen}
\affiliation{Department of Physics, P.O. Box 35 (YFL), FI-40014
University of Jyv\"askyl\"a, Finland}

\author{D. Tarpanov}
\affiliation{Institute of Theoretical Physics, Faculty of Physics,
University of Warsaw, ul. Ho\.za 69, PL-00-681 Warsaw, Poland}
\affiliation{Institute for Nuclear Research and Nuclear Energy, 1784
Sofia, Bulgaria}

\date{\today}

\begin{abstract}
Parameters of nuclear energy-density-functionals (EDFs) are always
derived by an optimization to experimental data. For the
minima of appropriately defined penalty functions, a statistical
sensitivity analysis provides the uncertainties of the EDF
parameters. To quantify theoretical errors of observables given by
the model, we studied the propagation of uncertainties within the
{\UNEDFZERO} Skyrme-EDF approach. We found that typically the
standard errors rapidly increase towards neutron rich nuclei. This
can be linked to large uncertainties of the isovector coupling
constants of the currently used EDFs.
\end{abstract}


\maketitle

{\it Introduction}.
The nuclear energy-density-functional (EDF) approach is the only
microscopic method that can be applied throughout the entire nuclear
chart. Owing to the existing and future radioactive-beam facilities,
the experimentally known region of the nuclear landscape will be
pushed towards more and more exotic systems. New experimental data on
nuclei with a large neutron or proton excess will put predictions of
the current and future EDF models to test. Therefore, it is important
to assess the uncertainties and predictive power of models
\cite{[Toi08],[Dud10],[Rei10a],[Erl12]}.

The nuclear EDF models can best be formulated in terms of effective
theories, tailored to low-energy phenomena with high-energy effects
taken into account through model parameter adjustments. Although a
rigorous application of the effective-theory
methodology~\cite{[Dob12a]} is not yet fully implemented, the aspect
of systematic adjustments of model parameters to data is always a
crucial element of the approach. Only very recently, such adjustments
are accompanied with the full covariance error
analysis~\cite{[Klu09],[Kor10b],[Fat11],[Kor12]}.

The aim of the present work is to study how uncertainties in the
model parameters propagate to observables calculated within the
model. This appears to be the simplest and most fundamental method to
assess the predictive power of the model. First, the model is
characterized not only by the fit residuals, that is, deviations
between measured and calculated observables, but also by propagated
uncertainties, which carry the errors of model parameters over to
calculated observables. In this way, even the data points that have
been used for the parameter adjustments are characterized by the
propagated uncertainties.

A properly balanced parameter adjustment should, in principle, lead
to propagated uncertainties, which are comparable in magnitude to fit
residuals. Second, extrapolations of the model to unmeasured data
points, which is, in fact, one of the principal goals of formulating
models, bear no meaning if they are not accompanied by their
propagated uncertainties. Indeed, large propagated uncertainties
simply mean that relatively small deviations of model parameters,
within the bounds of the adjustment, strongly influence the predicted
values. On the one hand, this may give us invaluable information
on which parameters of the model are grossly undetermined, and on the
other hand, this may indicate which observables, when measured, would
have the strongest impact on decreasing the errors of parameters.

There is, of course, another important class of theoretical
uncertainties, which is not directly accessible through the
statistical analysis of parameters, and which is very difficult to
estimate, namely, the one stemming from the definition and limitation
of the model itself, see discussion in Refs.~\cite{[Toi08],[Rei10a],[Erl12],[Pie12]}.
However, the statistical analysis may also here
give us interesting information. Indeed, in cases when the propagated
uncertainties turn out to be significantly smaller than the fit
residuals, one may suspect that the model lacks important physics
and/or new terms in the Hamiltonian (functional) and extended
parameter freedom. This is so, because we are then in the situation
of model parameters well constrained by some data, but not describing
some other data correctly. This can be contrasted with the situation
of propagated uncertainties larger than the fit residuals, when one
cannot tell whether the model is deficient or whether
its parameters are
underdetermined.

{\it Method}.
A good measure of theoretical uncertainty is to calculate the
statistical standard deviation of an observable, predicted by the
model. In the present, work we used the Skyrme EDFs~\cite{[Ben03]} within
the Hartree-Fock-Bogoliubov (HFB) nuclear structure calculations. To
determine the propagation of uncertainties from the model parameters to some
observable, predicted by the model, one needs information about the
standard deviations and correlations between the model parameters.
Some of the recent Skyrme-EDF parameter optimizations have provided
this information \cite{[Klu09],[Kor10b],[Kor12]}. In this work, to
determine the propagation of theoretical uncertainties, we have
chosen to use the {\UNEDFZERO} parameterization of the Skyrme
EDF~\cite{[Kor10b]}.

The calculated statistical error of an observable is linked with the covariance
matrix of the nuclear EDF model parameters.
From the Taylor expansion, we can find that the square of statistical error reads,
\begin{equation}
\sigma^{2}(y) = \sum_{i,j=1}^{n}
{\rm Cov}\left( x_i,x_j\right)
\left[\frac{\partial y}{\partial x_{i}}\right]
\left[\frac{\partial y}{\partial x_{j}}\right] \, ,
\label{eq:stderr}
\end{equation}
where $y$ is the calculated observable, $x_i$ is the parameter of the
EDF and ${\rm Cov}(x_i,x_j)$ is the covariance matrix between
parameters $x_{i}$ and $x_{j}$. In the present work we use the covariance
matrix that was obtained from the sensitivity analysis of the
{\UNEDFZERO} parameterization, see Tables VIII and X of
Ref.~\cite{[Kor10b]}. Another use of the covariance matrix consists
in calculating covariance ellipsoids between pairs of
observables~\cite{[Rei10a],[Fat11]} and analyzing correlations
between observables~\cite{[Fat12],[Rei13]}.

Calculation of the standard error (\ref{eq:stderr}) requires
calculation of derivatives of observable $y$ with respect to model
parameters $x_{i}$. As the relations between parameters and
observables can be complicated, to calculate the derivatives one has
to use numerical methods. By definition, the derivative of a
function $f$ can be approximated simply by a finite difference,
$f'(a)\approx{}G_{0}(h)=\left(f(a+h)-f(a-h)\right)/2h$.
To improve the accuracy, we used the Richardson extrapolation
method \cite{[Ric27]},
\begin{equation}
G_{m}(h)=\frac{4^{m}G_{m-1}(h/2)-G_{m-1}(h)}{4^{m}-1}\, ,
\end{equation}
with $m=1,2,\ldots$, which then satisfies,
$f'(a)-G_{m}(h)={\mathcal O}(h^{2(m+1)})$.
Generally speaking, a larger $m$ means improved accuracy on
derivatives with an increased cost of computational time. In this
work we extrapolate up to $m=2$ as a balanced point.

Large values $h$ may lead to inaccurate derivatives, whereas small
values may suffer from rounding errors. In this paper, we checked
that when $h$ changes for any parameter between 0.001\% and 0.1\% of
the parameter's value, the calculated errors of observables do not
change much. Thus values of $h$ fixed in this region were used.

The HFB calculations were carried out by using the computer code
{\HOSPHE}~\cite{[Car10d],[Car13]}, which solves the HFB equations
in a spherical harmonic-oscillator basis. We have run it within
exactly the same conditions as those defined for the optimization of
{\UNEDFZERO}. In particular, calculations were done in the space of
20 major oscillator shells and the Lipkin-Nogami procedure of
Ref.~\cite{[Sto03]} with the pairing cut-off window of $E_{\rm
cut}=60\,{\rm MeV}$ was used. We determined ground-state properties
of all even-even semi-magic nuclei with $N=20$, 28, 50, 82, and 126
and $Z=20$, 28, 50, and 82 extending between the two-proton and
two-neutron driplines.

For {\UNEDFZERO}, some of the very neutron rich Ca, Sn, and Pb
isotopes turn to be deformed, see, e.g., supplementary material of
Ref.~\cite{[Erl12]}. The same also holds for some semimagic isotonic
chains at the very neutron-rich regime. Since the current
calculations are done with a spherical HFB solver, the calculated
uncertainties may lack deformation effects. {\it A priori}, one can
expect that the standard errors determined in spherical and
deformed nuclei should be similar, see also Ref.~\cite{[Kor13]}, although in the future this
conjecture must be tested in more extensive calculations. The aim of
this work is to quantify a typical magnitude of the standard
error related to an observable, and thus assess the overall
predictive power of the model.

\begin{figure}[tbh]
\center
\includegraphics[width=0.95\columnwidth]{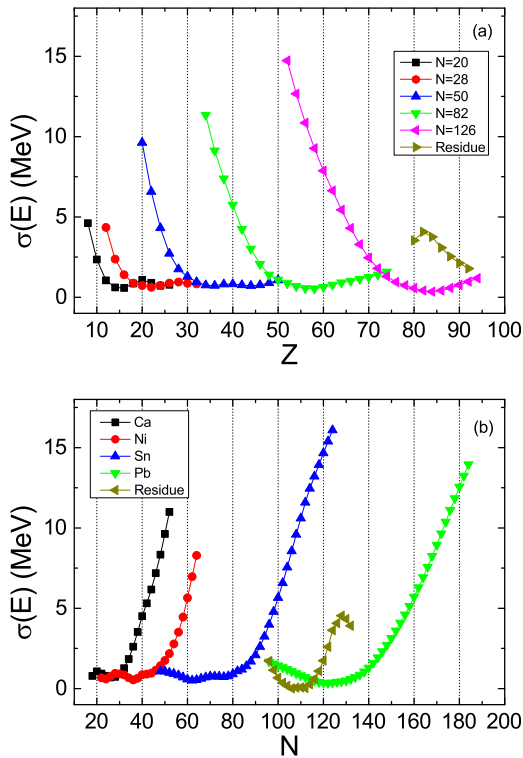}
\caption[Fig 1.]{(Color online) Standard errors of binding energies
of semi-magic nuclei, calculated for (a) isotonic chains with magic
neutron numbers and (b) isotopic chains with magic proton numbers.
For the Pb and $N=126$ nuclei, moduli of fit residuals with respect to
experimental data \protect\cite{[Aud12]} are also shown.}
\label{fig:stderrorBpn}
\end{figure}

{\it Results}.
In Fig.~\ref{fig:stderrorBpn}, we present the calculated standard
errors of binding energies of semi-magic nuclei. The upper panel
shows the isotonic chains with magic neutron numbers of $N=20$, 28,
50, 82, and 126 and the lower panel shows the Ca, Ni, Sn, and Pb
isotopic chains. Within the experimentally known regions of nuclei,
where also data points were used during the EDF optimization,
standard error usually stays rather constant and is typically of the
order of 1\,MeV. Moving to the neutron rich regime, calculated standard
error increases rather steadily. This is directly related to the fact
that the isovector parameters of the EDF are not constrained as
well as the isoscalar ones~\cite{[Rei10a],[Kor10b]}.
With isotonic (isotopic) chains, the
nuclei with small $Z$ (large $N$) of Fig.~\ref{fig:stderrorBpn} are
the ones with most neutron excess, and therefore standard deviation
is largest in these nuclei. The values reached there can be of the
order of 10\,MeV, which means that for the total binding energies,
when approaching the neutron drip line the predictive power of the
model is rather poor.

For the Pb and $N=126$ nuclei we also show, as an illustration, the
moduli of fit residuals with respect to experimental data
\protect\cite{[Aud12]}. We note immediately, that the propagated
error of the binding energy of $^{208}$Pb of 0.41\,MeV is much
smaller than the fit residual of 4.07\,MeV. This clearly
indicates that no refined fits can probably account for this
experimental data point, because the fit parameters are already quite
rigidly constrained by other data. This result points to a missing
physics in the description of the $^{208}$Pb mass, which is most
probably related to a poor description of the ground-states
collective correlations in doubly magic systems, see discussion in
Refs.~\cite{[Ben06],[Del10],[Car13a]}.

\begin{figure}[tbh]
\center
\includegraphics[width=0.95\columnwidth]{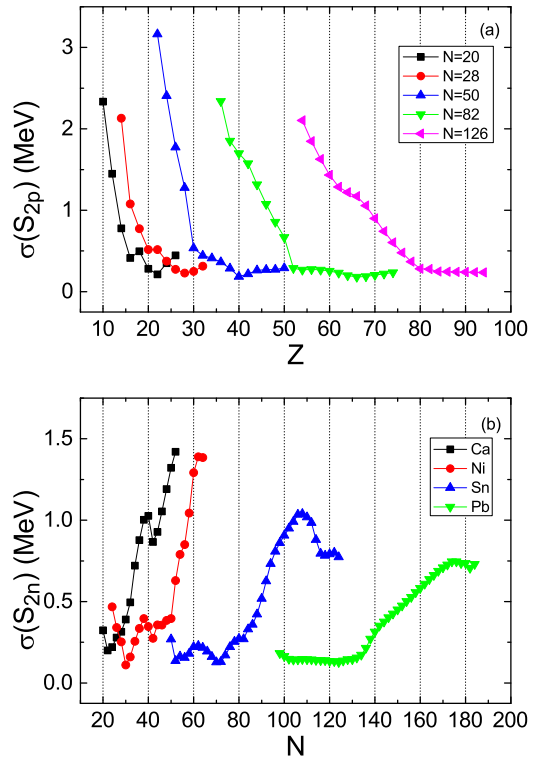}
\caption[Fig 2.]{(Color online) (a) Standard errors of two-proton
separation energies $S_{2{\rm p}}$ for neutron semi-magic
nuclei. (b) Standard errors of two-neutron separation energies
$S_{2{\rm n}}$ for proton semi-magic nuclei.}
\label{fig:stderrorS2NP}
\end{figure}

In Fig.~\ref{fig:stderrorS2NP}, the standard errors of two-neutron
($S_{2{\rm n}}$) and two-proton ($S_{2{\rm p}}$) separation energies
are shown. Similarly as for the binding energies, within the
experimentally know region, these standard errors stay rather small,
that is, in heavy nuclei below 0.5\,MeV for $S_{2{\rm p}}$ and below
0.25\,MeV for $S_{2{\rm n}}$. Then, they start to increase rapidly
when the calculations are extrapolated to neutron-rich regime,
especially after the doubly magic gaps are crossed. Owing to the
cancellation of errors, even in very exotic nuclei, the standard
errors of two-particle separation energies are much smaller than
those of the binding energies. This nicely illustrates the robustness
of the position of two-neutron driplines discussed in
Ref.~\cite{[Erl12]}.

\begin{figure}[tbh]
\center
\includegraphics[width=0.95\columnwidth]{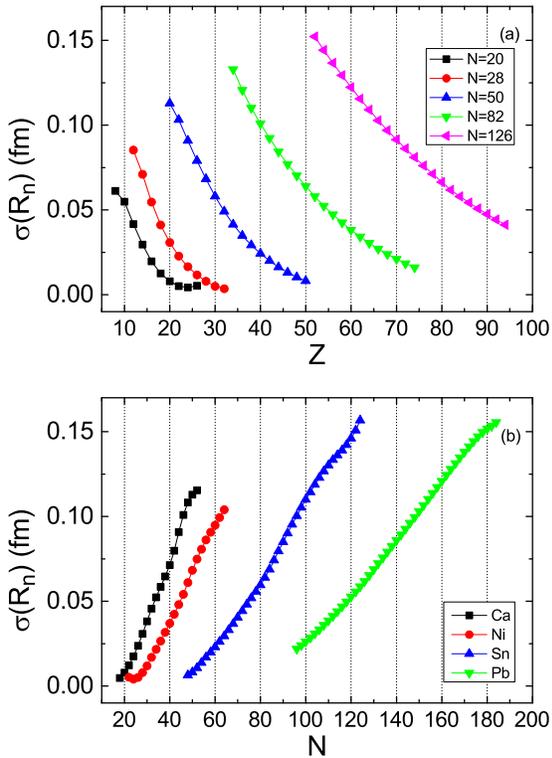}
\caption[Fig 3.]{(Color online) Same as in
Fig.~\protect\ref{fig:stderrorBpn} but for the neutron rms radii.}
\label{fig:stderrorRadN}
\end{figure}

\begin{figure}[tbh]
\center
\includegraphics[width=0.95\columnwidth]{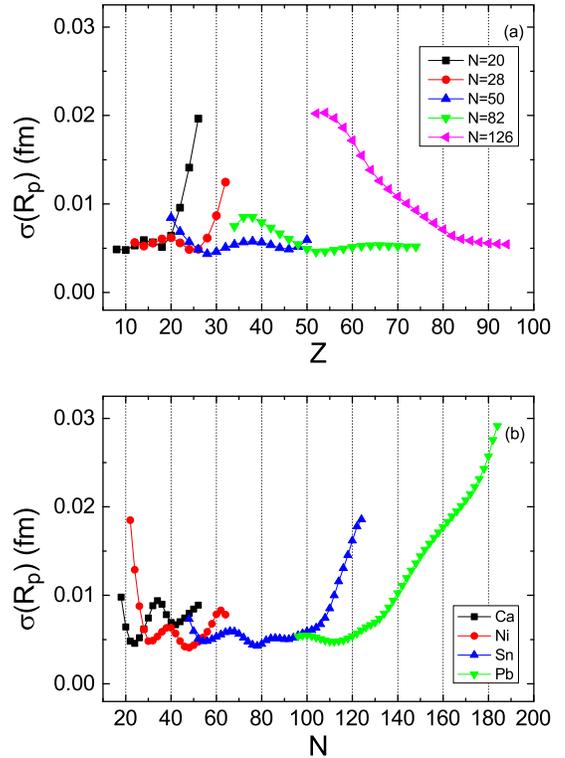}
\caption[Fig 2.]{(Color online)  Same as in
Fig.~\protect\ref{fig:stderrorBpn} but for the proton rms radii.}
\label{fig:stderrorRadP}
\end{figure}

In Figs.~\ref{fig:stderrorRadN} and \ref{fig:stderrorRadP}, we plot
the standard errors of neutron and proton rms radii, respectively.
One can see that the uncertainties related to the proton rms radii
are much smaller than those corresponding to neutron rms radii. This
is directly related to the fact that in the optimization of the
{\UNEDFZERO} parameter set, experimental input on the charge radii
was used. For example, in $^{208}$Pb we obtain $\sigma(R_{\rm
n})$=0.062\,fm and $\sigma(R_{\rm p})$=0.006\,fm. The
standard errors of neutron radii steadily increase towards the
neutron drip line, again illustrating the large uncertainties in the
isovector coupling constants of the functional. In heavy nuclei, the
same is also true for protons, however, in light nuclei, the
standard errors of proton radii also increase towards the proton drip
line. This effect can be related to the weak binding of protons
within a low Coulomb barrier. Uncertainties related to the proton and
neutron rms radii are closely linked to the uncertainties of the
neutron skin, see discussion in Ref.~\cite{[Kor13]}.

\begin{figure}[tbh]
\center
\includegraphics[width=0.95\columnwidth]{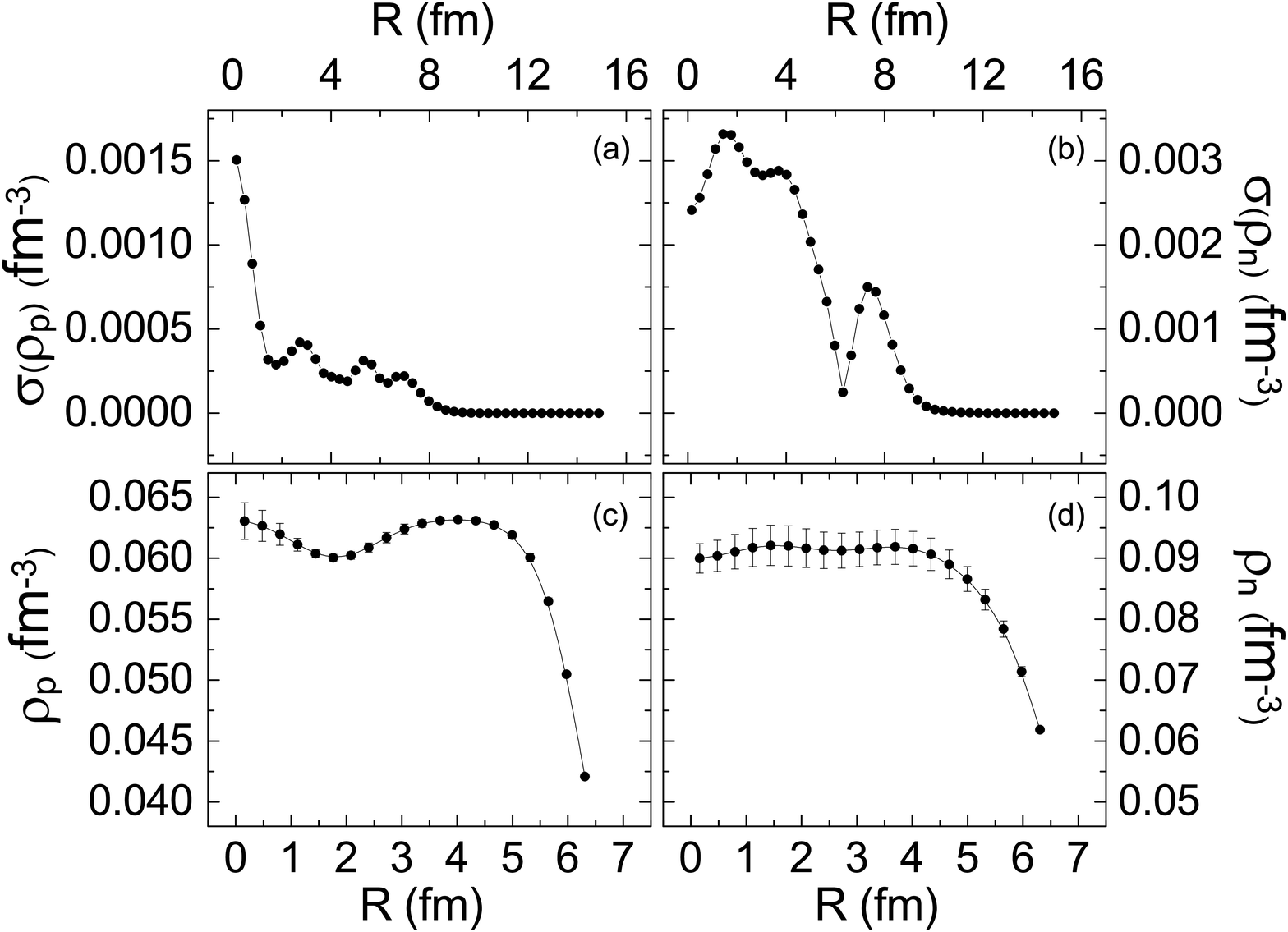}
\caption{Standard errors of densities in $^{208}$Pb for (a) protons
and (b) neutrons as functions of the radial variable $R$.
Panels (c) and (d) show the proton and neutron densities,
respectively, with the standard errors represented by error bars.}
\label{fig:stderrorDensity}
\end{figure}

In Fig.~\ref{fig:stderrorDensity}, we present standard errors of the
calculated mean-field (point-particle) proton and neutron densities
in $^{208}$Pb. Again we see that the standard errors of proton
densities are significantly smaller than for neutrons, and that the
neutron ones extend to much higher distances. The neutron density of
Fig.~\ref{fig:stderrorDensity}d can be compared with the experimental
weak-charge profile presented in Fig.~1 of Ref.~\cite{[Hor12]},
obtained from the PREX measurement at Jefferson Lab \cite{[Abr12]}.
In the region of relative flat neutron density profile (that is,
$R<4\,{\rm fm}$) we find our standard errors to be $0.0035\,{\rm
fm}^{-3}$ or less. In the same region, the incoherent sum of
experimental and model errors \cite{[Hor12]} is roughly
$0.0075\,{\rm fm}^{-3}$. Even though the density profile of
Fig.~\ref{fig:stderrorDensity} is for a point-neutron density,
folding it with a finite weak-charge distribution of neutrons, or
including in the experimental estimate other small corrections,
should not significantly change our result of propagated
theoretical uncertainties being about twice smaller than the
experimental ones.

\begin{figure}[tbh]
\center
\includegraphics[width=0.50\columnwidth]{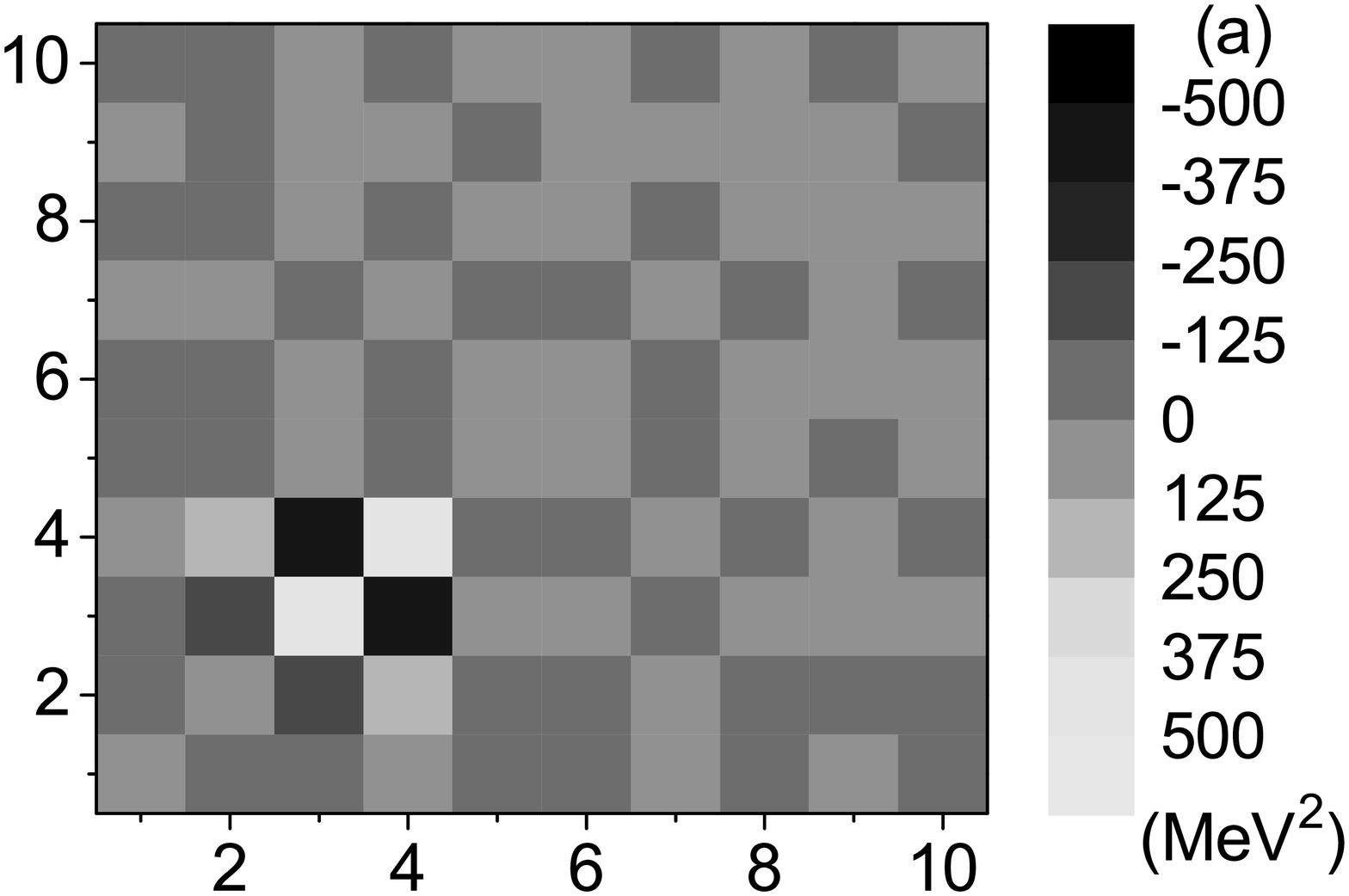}%
\includegraphics[width=0.50\columnwidth]{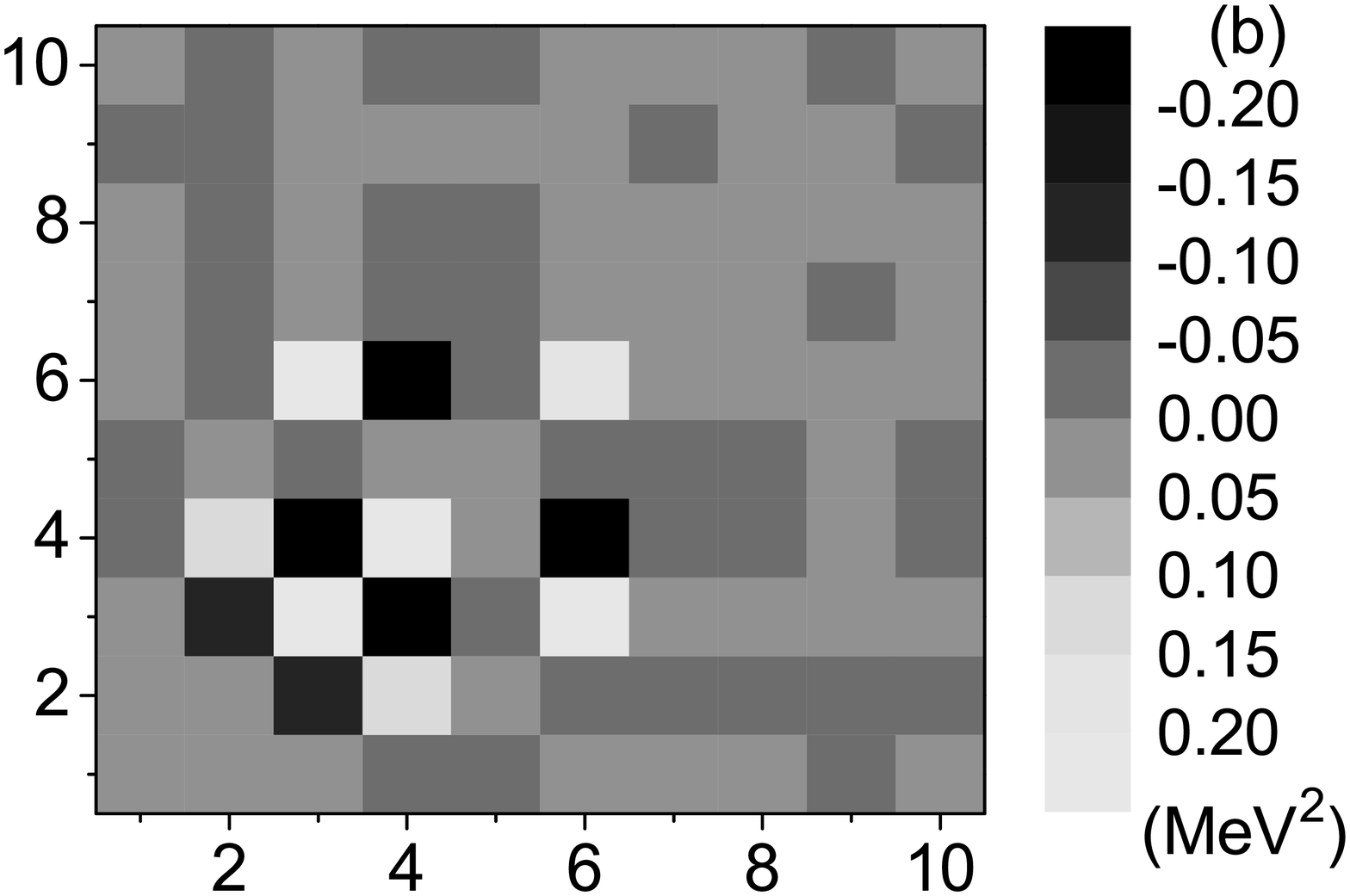}\\[2ex]
\includegraphics[width=0.50\columnwidth]{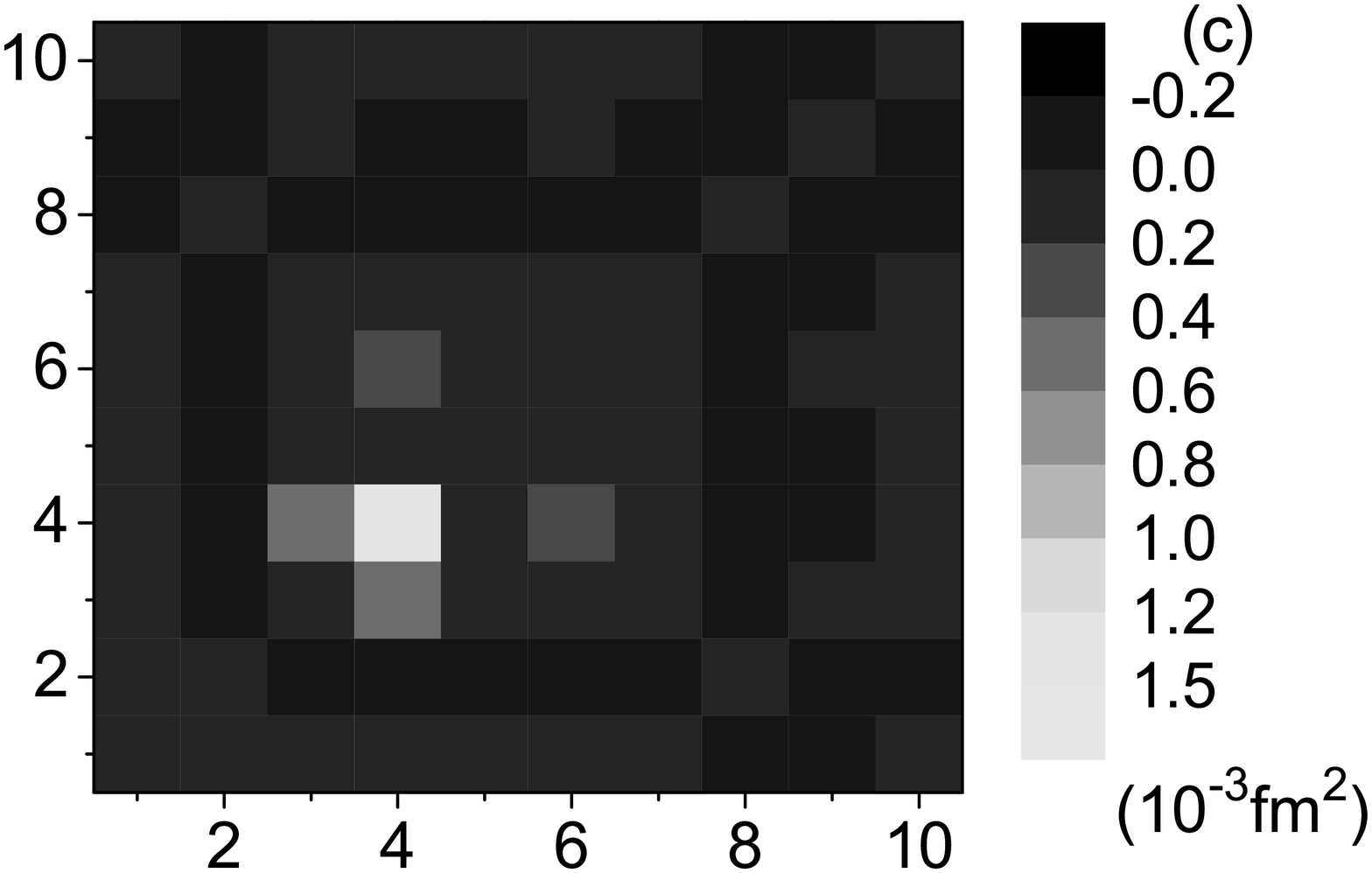}%
\includegraphics[width=0.50\columnwidth]{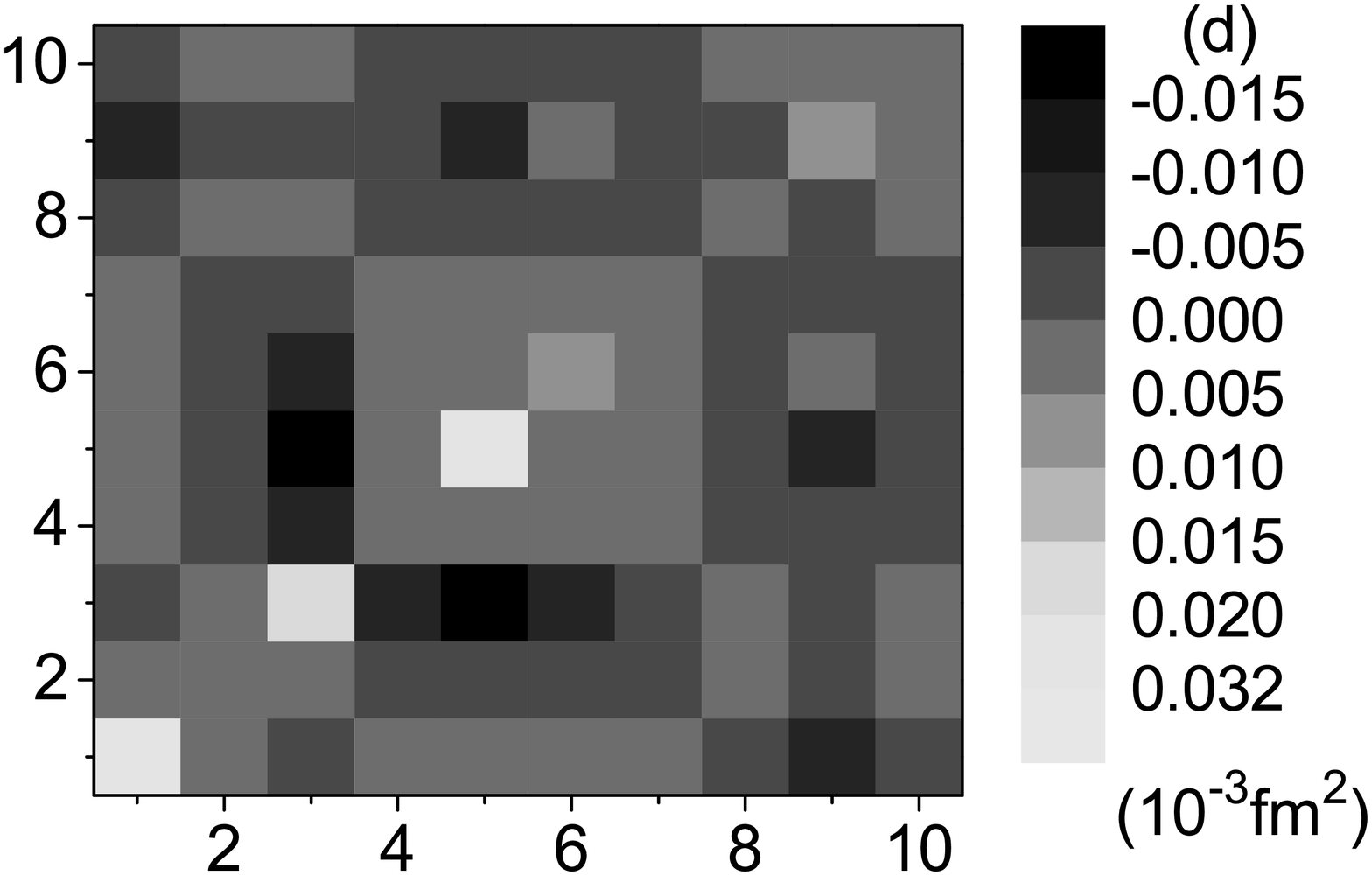}
\caption{Contributions of various components of the
sum~(\protect\ref{eq:stderr}) to squares of standard errors of the binding
energy $E$ (a), two-neutron separation energy $S_{2\rm n}$ (b),
neutron rms radius $R_{\rm n}$ (c), and proton rms radius $R_{\rm p}$ (d)
in $^{208}$Pb. The numbers given in the ordinates and abscissas refer
to the order of parameters defined in Ref.~\protect\cite{[Kor10b]}, that is,
$\rhoc$,
$\enm/A$,
$\asym$,
$\lsym$,
$C_{0}^{\rho\Delta\rho}$,
$C_{1}^{\rho\Delta\rho}$,
$V_0^n$,
$V_0^p$,
$C_{0}^{\rho\nabla J}$, and
$C_{1}^{\rho\nabla J}$.
}
\label{fig:S2Ncontributions}
\end{figure}

Owing to the correlations between the model parameters, the standard
errors calculated by using Eq.~(\ref{eq:stderr}) can be strongly
affected by the off-diagonal components of the covariance matrix.
This is illustrated in Fig.~\ref{fig:S2Ncontributions}, where various
components of the sum contributing to squares of standard errors of the binding
energy $E$ (a), two-neutron separation energy $S_{2\rm n}$ (b),
neutron rms radius $R_{\rm n}$ (c), and proton rms radius $R_{\rm p}$ (d)
in $^{208}$Pb are plotted.
The order of the parameters used in Fig.~\ref{fig:S2Ncontributions} is the same
as that used in the definition of the correlation matrix of {\UNEDFZERO}, see Tables VIII and X of
Ref.~\cite{[Kor10b]}. For the sake of completeness, these parameters are
also listed in the caption of Fig.~\ref{fig:S2Ncontributions}.

One sees from Fig.~\ref{fig:S2Ncontributions}a that the standard
error of the binding energy $E$ results from a strong cancellation
between large positive ($>500$\,MeV$^2$) diagonal contributions
coming from $\asym $ and $\lsym $ (parameters No.~3 and 4), and large
negative ($<-500$\,MeV$^2$) contributions coming from the
off-diagonal correlations between these two parameters. The remaining
parameters of the functional give much smaller positive and negative
contributions, bringing the final value to
$\sigma^2(E)$=0.17\,MeV$^2$. It is interesting to note that even in
this well-bound nucleus, the largest uncertainties of the binding
energy come from the poor symmetry-energy characterization of the
functional.

The situation obtained for the standard error of $S_{2\rm n}$
(Fig.~\ref{fig:S2Ncontributions}b) is quite similar. Although the
value of $\sigma^2(S_{2\rm n})$=0.02\,MeV$^2$ is much smaller than
that for the binding energy, it results form the cancellation of
contributions coming form the above two parameters and
$C_{1}^{\rho\Delta\rho}$ (parameters No.~6), which is the parameter
characterizing the isovector surface properties of the functional.
Again, the uncertainties in the isovector characterization of the
functional dominate the standard error here.

The pattern obtained for the standard error of the neutron rms radius
of $\sigma^2(R_{\rm n})$=3.8$\times$10$^{-3}$\,fm$^2$
(Fig.~\ref{fig:S2Ncontributions}c) is markedly different, with the
single parameter $\lsym $ dominating the uncertainty of this
variable, although again the isovector properties are to be blamed.
Only for the proton rms radius (Fig.~\ref{fig:S2Ncontributions}d),
its fairly small standard error of $\sigma^2(R_{\rm
p})$=4.1$\times$10$^{-5}$\,fm$^2$ is given by contributions coming
from $\asym$ and isoscalar parameters $\rhoc$ and
$C_{0}^{\rho\Delta\rho}$.

Even though the single-particle (s.p.) energies are not
strictly-speaking observables, they are nevertheless closely
connected to observable levels of neighboring even-odd nuclei and to
other nuclear properties like, e.g., deformations of open-shell
systems. It is, therefore, interesting to discuss their standard
errors and check how tightly their values are fixed by the adjustment
of the functional to other observables, see also discussion in Ref.~\cite{[Klu09]}.
Since our calculations
were performed with pairing correlations and Lipkin-Nogami
corrections taken into account, we performed the analysis
for the s.p.\ energies $e_{\sub{HF}}$ defined as the eigenstates
of the mean-field part of the HFB Hamiltonian.

In Table~\ref{table:spe}, we listed the s.p.\ energies and their
standard errors calculated in $^{208}$Pb. We see that the proton
(neutron) rms of standard errors, equal to 0.196 (0.176)\,MeV, is
significantly smaller than the rms deviations with respect to
empirical data~\cite{[Sch07]} of 0.52 (0.61)\,MeV. The rms
deviations obtained in lead are still much smaller than those
obtained when all doubly-magic nuclei are considered simultaneously,
which typically equal to about 1.2\,MeV~\cite{[Kor08]}, quite
independently of which functional and which data evaluation is used.
Anyhow, we see again that the s.p.\ energies, which in the fit of the
{\UNEDFZERO} functional were not taken into account, are rather
tightly constrained by the fit to other observables. It is,
therefore, unlikely that, within the parameter space of the standard
Skyrme functional, their agreement with data can be improved, which
supports conclusions reached in Ref.~\cite{[Kor08]}.

\begin{table}[tbh]
\begin{center}
\caption{The $^{208}$Pb proton (top) and neutron (bottom) s.p.\ energies $e_{\sub{HF}}$ (b) and
their standard errors (c), as compared
to the empirical values (d)~\protect\cite{[Sch07]},
residuals, $\Delta(e_{\sub{HF}})=e_{\sub{HF}}-e_{\text{exp}}$ (e),
PVCs $\delta e_{\sub{PVC}}$  (f), and
residuals of the PVC-corrected s.p.\ energies
$\Delta(e_{\sub{PVC}})=e_{\sub{HF}}+\delta e_{\sub{PVC}}-e_{\text{exp}}$ (g).
Where applicable, we also give the root-mean-square (rms) values
of entries shown in a given column. All energies are in MeV.}
\label{table:spe}
\begin{ruledtabular}
\begin{tabular}{lrrrrrr}
\multicolumn{1}{c}{orbital                  }&
\multicolumn{1}{c}{$e_{\sub{HF}}$           }&
\multicolumn{1}{c}{$\sigma(e_{\sub{HF}})$   }&
\multicolumn{1}{c}{$e_{\text{exp}}$         }&
\multicolumn{1}{c}{$\Delta(e_{\sub{HF}})$   }&
\multicolumn{1}{c}{$\delta e_{\sub{PVC}} $  }&
\multicolumn{1}{c}{$\Delta(e_{\sub{PVC}}$)  }\\
\multicolumn{1}{c}{(a)}  &
\multicolumn{1}{c}{(b)}  &
\multicolumn{1}{c}{(c)}  &
\multicolumn{1}{c}{(d)}  &
\multicolumn{1}{c}{(e)}  &
\multicolumn{1}{c}{(f)}  &
\multicolumn{1}{c}{(g)}  \\
\hline
$\pi$3p$_{1 /2}$ &  $ $0.219 & 0.181 & $-$0.16 & $ $0.38  & $-$0.440 & $-$0.06  \\
$\pi$3p$_{3 /2}$ &  $-$1.045 & 0.139 & $-$0.68 & $-$0.37  & $-$0.662 & $-$1.03  \\
$\pi$2f$_{5 /2}$ &  $-$1.284 & 0.229 & $-$0.97 & $-$0.31  & $-$0.480 & $-$0.79  \\
$\pi$1i$_{13/2}$ &  $-$2.794 & 0.238 & $-$2.10 & $-$0.69  & $-$0.221 & $-$0.92  \\
$\pi$1h$_{9 /2}$ &  $-$3.501 & 0.282 & $-$3.80 & $ $0.30  & $-$0.280 & $ $0.02  \\
$\pi$2f$_{7 /2}$ &  $-$3.725 & 0.115 & $-$2.90 & $-$0.83  & $-$0.284 & $-$1.11  \\
$\pi$3s$_{1 /2}$ &  $-$8.036 & 0.140 & $-$8.01 & $-$0.03  & $-$0.108 & $-$0.13  \\
$\pi$2d$_{3 /2}$ &  $-$8.378 & 0.199 & $-$8.36 & $-$0.02  & $ $0.220 & $ $0.20  \\
$\pi$1h$_{11/2}$ &  $-$9.153 & 0.207 & $-$9.36 & $ $0.21  & $-$0.141 & $ $0.07  \\
$\pi$2d$_{5 /2}$ & $-$10.117 & 0.117 & $-$9.82 & $-$0.30  & $ $0.116 & $-$0.18  \\
$\pi$1g$_{7 /2}$ & $-$10.908 & 0.229 &$-$12.00 & $ $1.09  & $ $0.131 & $ $1.22  \\
\hline
rms              &  n.a.     & 0.196 &  n.a.   &    0.52  &    0.328 &    0.70  \\
\hline
$\nu$3d$_{3 /2}$ &  $-$1.856 & 0.250 & $-$1.40 & $-$0.46  & $-$0.104 & $-$0.56  \\
$\nu$4s$_{1 /2}$ &  $-$2.051 & 0.235 & $-$1.90 & $-$0.15  & $-$0.668 & $-$0.82  \\
$\nu$2g$_{7 /2}$ &  $-$2.141 & 0.258 & $-$1.44 & $-$0.70  & $-$0.280 & $-$0.98  \\
$\nu$1f$_{15/2}$ &  $-$2.231 & 0.167 & $-$2.51 & $ $0.28  & $-$0.226 & $ $0.05  \\
$\nu$3d$_{5 /2}$ &  $-$2.549 & 0.166 & $-$2.37 & $-$0.18  & $-$0.384 & $-$0.56  \\
$\nu$1i$_{11/2}$ &  $-$2.680 & 0.223 & $-$3.16 & $ $0.48  & $-$0.271 & $ $0.21  \\
$\nu$2g$_{9 /2}$ &  $-$4.336 & 0.071 & $-$3.94 & $-$0.40  & $-$0.183 & $-$0.58  \\
$\nu$3p$_{1 /2}$ &  $-$7.855 & 0.109 & $-$7.37 & $-$0.49  & $ $0.152 & $-$0.33  \\
$\nu$3p$_{3 /2}$ &  $-$8.503 & 0.065 & $-$8.26 & $-$0.24  & $-$0.119 & $-$0.36  \\
$\nu$2f$_{5 /2}$ &  $-$8.519 & 0.143 & $-$7.94 & $-$0.58  & $ $0.156 & $-$0.42  \\
$\nu$1i$_{13/2}$ &  $-$8.603 & 0.162 & $-$9.24 & $ $0.64  & $-$0.130 & $ $0.51  \\
$\nu$1h$_{9 /2}$ &  $-$9.922 & 0.190 &$-$11.40 & $ $1.48  & $ $0.120 & $ $1.60  \\
$\nu$2f$_{7 /2}$ & $-$10.407 & 0.103 & $-$9.81 & $-$0.60  & $ $0.179 & $-$0.42  \\
\hline
rms              &  n.a.     & 0.176 &  n.a.   &    0.61  &    0.273 &    0.68
\end{tabular}
\end{ruledtabular}
\end{center}
\end{table}
Of course, in principle, the obtained insufficient agreement with
data can also signal some important missing physics. For the s.p.\
energies, an obvious and often invoked effect is the well-known
particle-vibration-coupling (PVC)~\cite{[Boh75]}, which continues to be
the subject of recent works~\cite{[Lit07],[Col10],[Miz12]}. Before studying this
problem for the Skyrme functionals in full detail~\cite{[Tar13]}, in
Table~\ref{table:spe} we presented the PVC
corrections to s.p.\ energies, $\delta e_{\sub{PVC}}$, calculated
with the standard methodology reviewed recently in, e.g.,
Ref.~\cite{[Col10]}. We see that in the particular case studied here,
the inclusion of the PVCs does not improve the agreement with data,
but, in fact, the rms deviation obtained for protons (neutrons)
even increases to 0.70 (0.68)\,MeV. An improved agreement with data
should probably be sought for by using richer parameter spaces
of the functional, such as, e.g., those proposed in Refs.~\cite{[Car08e],[Dob12a]}.
Evidently, further studies of this problem are very much required.

{\it Summary and Conclusions}.
In the framework of the {\UNEDFZERO} Skyrme-EDF model, we
studied propagation of uncertainties from model parameters to
calculated observables. We found that the magnitude of standard
errors increases towards neutron rich nuclei. This is directly linked
to the isovector part of the functional, where coupling constants
have large uncertainties.

For binding energies, the standard errors are in the range of one to
several MeV -- towards the neutron drip line rapidly increasing to
about 10\,Mev. For two-particle separation energies, they range from
a few hundred keV to a few MeV. Compared to the deviations from
experimental values, the standard errors of binding energies are, in
the experimentally known region, of the same order of magnitude or
somewhat smaller. For two-particle separation energy, standard errors
are typically smaller, but still comparable. It is worth stressing
that the theoretical uncertainties of binding energies and
two-particle separation energies are still much larger than the
present experimental errors in Penning-trap
measurements~\cite{[Hak12],[Kan12]}. For proton rms radii, our
standard errors are also comparable, but slightly smaller than the
rms deviations from experimental values. Theoretical uncertainties
related to the neutron skin are actually smaller than the current
experimental estimates~\cite{[Kor13]}.

In conclusion, determination of theoretical uncertainties in terms of
standard statistical errors of observables provides a way to assess
the predictive power of nuclear models. In the future EDF parameter
optimizations, it would be useful to perform by default the
sensitivity analyses of the obtained minima. With this information,
theoretical uncertainties related to the model parameters and
predictions should be routinely assessed.

{\it Acknowledgments}.
Interesting comments by Witek Nazarewicz are gratefully acknowledged.
We thank Gillis Carlsson for his help with upgrading the numerical code.
This work was supported in part by the Academy of Finland and
University of Jyv\"askyl\"a within the FIDIPRO programme, by the
Centre of Excellence Programme 2012--2017 (Nuclear and Accelerator
Based Physics Programme at JYFL), and by the European Union's Seventh
Framework Programme ENSAR (THEXO) under Grant No.\ 262010. We
acknowledge the CSC - IT Center for Science Ltd, Finland, for the
allocation of computational resources.


\begin{thebibliography}{10}

\bibitem{[Toi08]}
{J. Toivanen, J. Dobaczewski, M. Kortelainen, and K. Mizuyama, Phys. Rev. C
  {\bf 78}, 034306 (2008)}.

\bibitem{[Dud10]}
{J. Dudek, B. Szpak, M.-G. Porquet, H. Molique, K. Rybak, and B. Fornal, J.
  Phys. G {\bf 37}, 064031 (2010)}.

\bibitem{[Rei10a]}
{P.-G. Reinhard and W. Nazarewicz, Phys. Rev. C {\bf 81}, 051303 (2010)}.

\bibitem{[Erl12]}
{J. Erler, N. Birge, M. Kortelainen, W. Nazarewicz, E. Olsen, A. M. Perhac, and
  M. Stoitsov, Nature {\bf 486}, 509 (2012)}.

\bibitem{[Dob12a]}
{J. Dobaczewski, K. Bennaceur, and F. Raimondi, J. Phys. G {\bf 39}, 125103
  (2012)}.

\bibitem{[Klu09]}
{P. Kl\"upfel, P.-G. Reinhard, T.J. Burvenich, and J.A. Maruhn, Phys. Rev. C
  {\bf 79}, 034310 (2009)}.

\bibitem{[Kor10b]}
{M. Kortelainen, T. Lesinski, J. Mor\'{e}, W. Nazarewicz, J. Sarich, N.
  Schunck, M.V. Stoitsov, and S. Wild, Phys. Rev. C {\bf 82}, 024313 (2010)}.

\bibitem{[Fat11]}
{F.J. Fattoyev and J. Piekarewicz, Phys. Rev. C {\bf 84}, 064302 (2011)}.

\bibitem{[Kor12]}
{M. Kortelainen, J. McDonnell, W. Nazarewicz, P.-G. Reinhard, J. Sarich, N.
  Schunck, M.V. Stoitsov, and S. Wild, Phys. Rev. C {\bf 85}, 024304 (2012)}.

\bibitem{[Pie12]}
{J. Piekarewicz, B.K. Agrawal, G. Col\`o, W. Nazarewicz, N. Paar, P.-G.
  Reinhard, X. Roca-Maza, and D. Vretenar, Phys. Rev. C {\bf 85}, 041302(R)
  (2012)}.

\bibitem{[Ben03]}
{M. Bender, P.-H. Heenen, and P.-G. Reinhard, Rev. Mod. Phys. {\bf 75}, 121
  (2003)}.

\bibitem{[Fat12]}
{F.J. Fattoyev and J. Piekarewicz, Phys. Rev. C {\bf 86}, 015802 (2012)}.

\bibitem{[Rei13]}
{P.-G. Reinhard and W. Nazarewicz, Phys. Rev. C {\bf 87}, 014324 (2013)}.

\bibitem{[Ric27]}
{L.F. Richardson and J.A. Gaunt, Phil. Trans. R. Soc. Lond. A {\bf 226} 299
  (1927)}.

\bibitem{[Car10d]}
{B.G. Carlsson, J. Dobaczewski, J. Toivanen, and P. Vesel\'y, Comput. Phys.
  Commun. {\bf 181}, 1641 (2010)}.

\bibitem{[Car13]}
{B.G. Carlsson, J. Toivanen, P. Vesel\'y, and Y. Gao, {\it unpublished}}.

\bibitem{[Sto03]}
{M.V. Stoitsov, J. Dobaczewski, W. Nazarewicz, S. Pittel, and D.J. Dean, Phys.
  Rev. C {\bf 68}, 054312 (2003)}.

\bibitem{[Kor13]}
{M. Kortelainen, J. Erler, N. Birge, Y. Gao, W. Nazarewicz, and E. Olsen, {\it
  unpublished}}.

\bibitem{[Aud12]}
{G. Audi, F.G. Kondev, M. Wang, B. Pfeiffer, X. Sun, J. Blachot, and
  M.MacCormick, Chinese Physics {\bf C36}, 1157 (2012); G. Audi, M. Wang, A.H.
  Wapstra, F.G. Kondev, M. MacCormick, X. Xu, and B. Pfeiffer, Chinese Physics
  {\bf C36}, 1287 (2012); M. Wang, G. Audi, A.H. Wapstra, F.G. Kondev, M.
  MacCormick, X. Xu, and B. Pfeiffer, Chinese Physics {\bf C36}, 1603 (2012)}.

\bibitem{[Ben06]}
{M. Bender, G. Bertsch, and P.-H. Heenen, Phys. Rev. C {\bf 73}, 034322
  (2006)}.

\bibitem{[Del10]}
{J.-P. Delaroche, M. Girod, J. Libert, H. Goutte, S. Hilaire, S. P\'eru, N.
  Pillet, and G.F. Bertsch, Phys. Rev. C {\bf 81}, 014303 (2010)}.

\bibitem{[Car13a]}
{B.G. Carlsson, J. Toivanen, and U. von Barth, arXiv:1212.2050}.

\bibitem{[Hor12]}
{C.J. Horowitz, Z. Ahmed, C.-M. Jen, A. Rakhman, P.A. Souder, M.M. Dalton, N.
  Liyanage, K.D. Paschke, K. Saenboonruang, R. Silwal, G.B. Franklin, M.
  Friend, B. Quinn, K.S. Kumar, D. McNulty, L. Mercado, S. Riordan, J. Wexler,
  R.W. Michaels, and G.M. Urciuoli, Phys. Rev. C {\bf 85}, 032501 (2012)}.

\bibitem{[Abr12]}
{S. Abrahamyan, Z. Ahmed, H. Albataineh, {\it et. al.}, Phys. Rev. Lett. {\bf
  108}, 112502 (2012)}.

\bibitem{[Sch07]}
{N. Schwierz, I. Wiedenhover, and A. Volya, arXiv:0709.3525}.

\bibitem{[Kor08]}
{M. Kortelainen, J. Dobaczewski, K. Mizuyama, and J. Toivanen, Phys. Rev. C
  {\bf 77}, 064307 (2008)}.

\bibitem{[Boh75]}
{A. Bohr and B.R. Mottelson, {\sl Nuclear Structure} (Benjamin, New York,
  1975), Vol. II}.

\bibitem{[Lit07]}
{E. Litvinova, P. Ring, and V. Tselyaev, Phys. Rev. C {\bf 75}, 064308 (2007)}.

\bibitem{[Col10]}
{G. Col\`o, H. Sagawa, and P.-F. Bortignon, Phys. Rev. C {\bf 82}, 064307
  (2010)}.

\bibitem{[Miz12]}
{K. Mizuyama, G. Col\`o, and E. Vigezzi, Phys. Rev. C {\bf 86}, 034318 (2012)}.

\bibitem{[Tar13]}
{D. Tarpanov {\it et al.}, unpublished}.

\bibitem{[Car08e]}
{B.G. Carlsson, J. Dobaczewski, and M. Kortelainen, Phys. Rev. C {\bf 78},
  044326 (2008); {\bf 81}, 029904(E) (2010)}.

\bibitem{[Hak12]}
{J. Hakala, J. Dobaczewski, D. Gorelov, T. Eronen, A. Jokinen, A. Kankainen,
  V.S. Kolhinen, M. Kortelainen, I.D. Moore, H. Penttil\"a, S. Rinta-Antila, J.
  Rissanen, A. Saastamoinen, V. Sonnenschein, and J. \"Ayst\"o, Phys. Rev.
  Lett. {\bf 109}, 032501 (2012)}.

\bibitem{[Kan12]}
{A. Kankainen, J. \"Ayst\"o, and A. Jokinen, J. Phys. G {\bf 39}, 093101
  (2012)}.

\end{thebibliography}

\end{document}